\documentclass{article}

\newcommand{\bt}{\begin{tabular}{c}}
\newcommand{\et}{\end{tabular}}

\newcommand{\eb}{\ee\be } 
\newcommand{\ebp}{\rt.\ee\be\lt.} 
\newcommand{\bmat}{\lt ( \begin{array} }
\newcommand{\emat}{  \end{array} \rt )}

\newcommand{\oH}{{\ov H}}
\newcommand{\oP}{{\ov P}}

\newcommand{\oQ}{{\ov Q}}

\newcommand{\oR}{{\ov R}}

\newcommand{\oB}{{\ov B}}

\newcommand{\ovD}{{\ov D}}

\newcommand{\oN}{{\ov N}}
\newcommand{\oJ}{{\ov J}}
\newcommand{\cd}{{\cdot}}

\newcommand{\oE}{{\ov E}}

\newcommand{\oU}{{\ov U}}

\newcommand{\ob}{{\ov b}}

\newcommand{\ovv}{{\ov v}}

\newcommand{\oq}{{\ov \q}}
\newcommand{\oT}{{\ov T}}
\newcommand{\oG}{{\ov \G}}

\newcommand{\oK}{{\ov K}}

\newcommand{\ED}{\end{document}}

\newcommand{\oY}{{\ov Y}}
\newcommand{\og}{{\ov g}}
\newcommand{\oy}{{\ov \y}}

\newcommand{\oC}{{\ov C}}
\newcommand{\oL}{{\ov L}}

\newcommand{\A}{{\ov A}}

\renewcommand{\a}{\alpha}	
\renewcommand{\b}{\beta}
\newcommand{\g}{\gamma}
\renewcommand{\d}{\delta}
\newcommand{\e}{\epsilon}
\newcommand{\ve}{\varepsilon}

\newcommand{\q}{\theta}

\newcommand{\f}{\phi}

\newcommand{\y}{\psi}
\newcommand{\w}{\omega}
\newcommand{\G}{\Gamma}
\newcommand{\D}{\Delta}
	
\renewcommand{\L}{\Lambda}
\newcommand{\X}{\Xi}
	
\renewcommand{\S}{\Sigma}

\newcommand{\la}{\label}
\newcommand{\ci}{\cite}

\newcommand{\ds}{\documentstyle}	
\newcommand{\fr}{\frac}

\newcommand{\pa}{\partial}
\newcommand{\ov}{\overline}
\newcommand{\be}{\begin{equation}}
\newcommand{\ee}{\end{equation}}
\newcommand{\ba}{\begin{array}} 
\newcommand{\ea}{\end{array}}
\newcommand{\bea}{\begin{eqnarray}}
\newcommand{\eea}{\end{eqnarray}}
\newcommand{\ra}{\rightarrow}
\newcommand{\Ra}{\Rightarrow}

\newcommand{\lt}{\left}
\newcommand{\rt}{\right}

\newcommand{\ben}{\begin{enumerate}}
\newcommand{\een}{\end{enumerate}}
\newcommand{\bitem}{\begin{itemize}}
\newcommand{\eitem}{\end{itemize}}
\newcommand{\articlenumber}{}

\newcommand{\articletitle}{Some Composite Hadrons and Leptons \\
which induce Supersymmetry Breaking \\
in the Supersymmetric Standard Model: \\
Cybersusy III}

\begin{document}
\makeatletter	   
\renewcommand{\ps@plain}{%
\renewcommand{\@oddhead}{{\articlenumber  \hspace{1cm} }\hspace{1cm}  \hfil\textrm{\thepage}} 
\renewcommand{\@evenhead}{\@oddhead}
\renewcommand{\@oddfoot}{\textrm{\articlenumber \hspace{1cm} }\hspace{1cm} \hfil\textrm{\thepage}}
\renewcommand{\@evenfoot}{\@oddfoot}}
\makeatother    
\title{  \articletitle \\ \articlenumber}
\author{ J. A.  Dixon\footnote{jadix@telus.net}\\ Dixon Law Firm\footnote{Fax: (403) 266-1487} \\1020 Canadian Centre\\
833 - 4th Ave. S. W. \\ Calgary, Alberta \\ Canada T2P 3T5 }
\maketitle
\pagestyle{plain}
\Large

\abstract{\large  This is the third paper in  a series of four papers which introduce cybersusy, which is a new mechanism for supersymmetry breaking in the supersymmetric standard model (SSM).  In this paper we display some solutions to the constraint equations of  BRS cohomology in the  SSM.  In particular we discuss the leptonic dotspinor pseudosupermultiplets that were used in Cybersusy I to calculate leptonic supersymmetry breaking in the supersymmetric standard model.  We also introduce  examples of    hadronic dotspinor pseudosupermultiplets that will induce baryonic  supersymmetry breaking in the SSM  for the  baryons with charge Q=-1, and related supersymmetry partner baryons. Some interesting relationships between the peculiar structure of the SSM, the existence of solutions for the BRS constraints, and supersymmetry breaking using cybersusy are noted. }

\Large

\section{Introduction}

\la{wfggrgrgeruyjkkyu}

\subsection{Review of Cybersusy in the First Two Papers}

In  \ci{cybersusyI}, cybersusy was introduced as a method to calculate the mass matrices of the electrons and the neutrinos and their superpartners, after taking supersymmetry breaking into account, in the standard supersymmetric model (SSM).  

 The introductory paper \ci{cybersusyI} reviewed the basic method and discussed the spectrum of masses somewhat, but it left some detailed work for the subsequent papers. The main things that were left to explain were:
\ben
\item
the derivation and application of the nilpotent operator $\d$ that arises from the BRS-ZJ  identity after the integration of the auxiliary fields in the Wess Zumino model;
\la{qwewqqf1}
\item
the structure of the fundamental chiral dotted pseudosuperfield ${\hat \f}^i_{\dot \a} $ and the fundamental  chiral scalar pseudosuperfield ${\hat A}^i$ which can be used as building blocks to construct simple composite chiral dotted pseudosuperfields
such as  ${\hat \w}_{\dot \a_1 \cdots \dot \a_{n}} $;
\la{qwewqqf2}
\item
the constraint equations that must be satisfied by the simple composite chiral dotted pseudosuperfields ${\hat \w}_{\dot \a_1 \cdots \dot \a_{n}} $, in order for them to be in the cohomology space of $\d$;
\la{qwewqqf3}
\item
the details of some of the leptonic and hadronic simple 
composite chiral dotted pseudosuperfields  ${\hat \w}_{\dot \a_1 \cdots \dot \a_{n}}$, and a discussion of the solution of the constraints in the SSM;
\la{qwewqqf4}
\item
the details of the computation of the lepton masses;
\la{qwewqqf5}
 \la{qwewqqf6}
\een

Items \ref{qwewqqf1},  \ref{qwewqqf2} and  \ref{qwewqqf3} in the foregoing list were addressed in \ci{cybersusyII}.   
Item \ref{qwewqqf4} in the foregoing list will be  addressed in this paper.   
Item \ref{qwewqqf5} in the foregoing list is the subject of  \ci{cybersusyIV}.

\subsection{Notation for the Massless Supersymmetric Standard Model}
\la{fjdkflvjvksdvsl}

In  \ci{cybersusyI},  we claimed that the baryons, and interesting leptons, appeared in a semi-magical sort of way as solutions of the constraint equations for the composite dotspinor part of the BRS cohomology in the massless SSM.  Here we give the details. Our notation for the SSM was set out in \ci{cybersusyI} and we shall repeat the table of quantum numbers from there: 

\be
\begin{tabular}{|c|c|c
|c|c|c
|c|c|c|}
\hline
\multicolumn{8}
{|c|}{Table of the Chiral Superfields in the SSM
}
\\
\hline
\multicolumn{8}
{|c|}{ \bf Superstandard Model, Left ${\cal L}$
Fields}
\\
\hline
{\rm Field} & Y 
& {\rm SU(3)} 
& {\rm SU(2)} 
& {\rm F} 
& {\rm B} 
& {\rm L} 
& {\rm D} 
\\
\hline
$ L^{pi} $& -1 
& 1 & 2 
& 3
& 0
& 1
& 1
\\
\hline
$ Q^{cpi} $ & $\fr{1}{3}$ 
& 
3 &
2 &
3 &
 $\fr{1}{3}$
& 0
& 1
\\\hline
$J$
& 0 
& 1
& 1
& 1
& 0
& 0
& 1
\\
\hline
\multicolumn{8}
{|c|}{ \bf Superstandard Model, Right 
${\cal R}$
Fields}
\\
\hline
$P^{ p}$ & 2 
& 1
& 1
& 3
& 0
& -1
& 1
\\
\hline
$R^{ p}$ & 0 
& 1
& 1
& 3
& 0
& -1
& 1
\\

\hline
$T_c^{ p}$ & $-\fr{4}{3}$ 
& ${\ov 3}$ &
1 &
3 &
 $-\fr{1}{3}$
& 0
& 1
\\
\hline
$B_c^{ p}$ & $\fr{2}{3}$ 
& ${\ov 3}$ 
&
1 &
3 &
 $- \fr{1}{3}$
& 0
& 1
\\
\hline
$H^i$ 
& -1 
& 1
& 2
& 1
& 0
& 0
& 1
\\
\hline
$K^i$ 
& 1 
& 1
& 2
& 1
& 0
& 0
& 1
\\
\hline
\end{tabular}
\\
\la{qefewfqwefwrrr}
\ee
\vspace{.2cm}

We used the lepton solutions in \ci{cybersusyI} to calculate supersymmetry breaking for the leptons.  Here we can close the loop  by exhibiting the full form of the dotspinor superfields that we used there, and deriving the algebra we used there also.

We repeat the superspace potential for the SSM:
\vspace{.2cm}
\be
\begin{tabular}{|c|c|c
|c|c|c
|c|c|c}
\hline
\multicolumn{8}{|c|}{\bf Table
 \ref{fqwefweef1212}
}
\\
\hline
\multicolumn{8}{|c|}{\bf Superpotential for SuperStandard Model}
\\
\hline
\multicolumn{8}{|c|}{$
P_{{\rm SP}}   =
g \e_{ij} H^i K^j J
- g_{\rm J} m^2 J
$}
\\
\multicolumn{8}{|c|}{$
+
p_{p q} \e_{ij} L^{p i} H^j P^{ q} 
+
r_{p q} \e_{ij} L^{p i} K^j R^{ q}
$}\\
\multicolumn{8}{|c|}{$
+
t_{p  q} \e_{ij} Q^{c p i} K^j T_c^{ q}
+
b_{p  q} \e_{ij} Q^{c p i} H^j B_c^{ q}
$}
\\
\hline
\end{tabular}
\\
\la{fqwefweef1212}
\ee

\subsection{Form of $d_3$ for the massless SSM }

In \ci{cybersusyII}, we wrote down the general constraint equation for simple dotspinor generators and their complex conjugates.  It uses the expression 

\be
d_3  
= 
C_{\a} {\ov g}^{ijk} \A_j \A_k 
\y_{\a}^{i \dag}
+
 \oC_{\dot \a}  
{g}_{ijk} A^j A^k \oy_{i \dot \a}^{\dag}
\ee
This is equivalent to:
\be
d_3 
= C_{\a}  
\fr{\pa \; P_{{\rm SP}}^* }{\pa \A_i } 
\y_{\a}^{i \dag}
\eb
+ \oC_{\dot \a} 
\fr{\pa \; P_{{\rm SP}}}{\pa A^i } 
\oy_{i \dot \a}^{\dag}
\la{fgregeger}
\ee
So using the (\ref{fgregeger})
 part of $d_3$ for the massless SSM, we get:  
\[
d_3 
= \oC_{\dot \a} 
\lt \{
\fr{\pa \; P_{{\rm SP}} }{\pa K^i } 
\oy_{K i\dot \a}^{ \dag}
\rt.
\]
\[
+
\fr{\pa \; P_{{\rm SP}} }{\pa H^i } 
\oy_{H i \dot \a}^{ \dag}
+
\fr{\pa \; P_{{\rm SP}} }{\pa J } 
\oy_{J\dot \a}^{ \dag}
\]
\[
+
\fr{\pa \; P_{{\rm SP}} }{\pa L^{p i} } 
\oy_{L i \dot \a}^{p  \dag}
+
\fr{\pa \; P_{{\rm SP}} }{\pa   P^{ q}  } 
\oy_{P\dot \a}^{ \dag}
\]
\[
+
\fr{\pa \; P_{{\rm SP}} }{\pa   R^{ q}  } 
\oy_{R q\dot \a}^{q \dag}
+
\fr{\pa \; P_{{\rm SP}} }{\pa Q^{c p i} } 
\oy_{Q c p i\dot \a}^{ \dag}
\]
\[
\lt.
+
\fr{\pa \; P_{{\rm SP}} }{\pa    T_c^{ q}  } 
\oy_{T  q\dot \a}^{c  \dag}
+
\fr{\pa \; P_{{\rm SP}} }{\pa   B_c^{ q}  } 
\oy_{B q\dot \a}^{c \dag}
\rt \}
\]
\be
+ {\rm Complex\; Conjugate}
\la{qefwefwefwefopjkpjopj}
\ee
and working out the derivatives yields:

\be
d_3 
= \oC_{\dot \a} 
\lt \{
g \e_{ij} H^i K^j  
\oy_{J \dot \a}^{ \dag}
\ebp
+
\lt (
g \e_{ij} H^i J
+
r_{p q} \e_{ij} L^{p i}  R^{ q}
+
t_{p  q} \e_{ij} Q^{c p i}   T_c^{ q}
\rt )
\oy_{K j \dot \a}^{ \dag}
\ebp
+
\lt (
g \e_{ij}  K^j J
-
p_{p q} \e_{ij} L^{p j}   P^{ q} 
 -
b_{p  q} \e_{ij} Q^{c p j} B_c^{ q}
\rt )
\oy_{H i\dot \a}^{ \dag}
\ebp
+
\lt (
 p_{p q} \e_{ij}  H^j P^{ q} 
+ r_{p q} \e_{ij}  K^j R^{ q}
\rt )
\oy_{L p i \dot \a}^{ \dag}
\ebp
+
\lt (
 p_{p q} \e_{ij} L^{p i} H^j 
\rt )
\oy_{P q \dot \a}^{ \dag}
+
\lt (
 r_{p q} \e_{ij} L^{p i} K^j 
\rt )
\oy_{R q\dot \a}^{\dag}
\ebp
+
\lt (
t_{p  q} \e_{ij}   K^j T_c^{ q}
+b_{p  q} \e_{ij}   H^j B_c^{ q}
\rt )
\oy_{Q c p i\dot \a}^{ \dag}
\ebp
+
\lt (
t_{p  q} \e_{ij} Q^{c p i} K^j 
\rt )
\oy_{T q\dot \a}^{c \dag}
+
\lt (
b_{p  q} \e_{ij} Q^{c p i} H^j 
\rt )
\oy_{B q\dot \a}^{c \dag}
\rt \} 
\eb
+ {\rm Complex\; Conjugate}
\la{egrggwrtrhtrhbntr}
\ee

Now we can look for solutions of the constraint equations in the SSM, using this specific form for $d_3$.

\subsection{An example of a hadronic dotspinor from the cohomology space}

\la{qwergregqergerger}

As announced above, here is an example of a baryon-type solution for the constraint equations:

\be
\w_{\dot \a}
=
f^{p_1}_{ p_2 p_3}
\ve^{c_1 c_2 c_3}
\lt \{
g  
\oy_{Q c_1 p_1 i \dot \a} K^i
+
b_{p_1 q}
\oy_{J\dot \a} \; 
B_{c_1}^{q}
\rt \}
B_{c_2}^{p_2} B_{c_3}^{p_3}
\la{wfgreerhjtuyj}
\ee
This is an example of a simple generator as discussed   for simple dotspinors in \ci{cybersusyII}.  Note that it has baryon number B=1, hypercharge Y=2, and that it will give rise to an antichiral undotted spinor pseudosuperfield with spin $J=\fr{1}{2}$, provided that the constraint is satisfied.

Its complex conjugate is
\be
\ov \w_{\a}
=
f_{p_1}^{ p_2 p_3}
\ve_{c_1 c_2 c_3}
\lt \{
\og 
\y^{c_1 p_1}_{Q\a} \cd \oK \; 
+
\ob^{p_1 q}
\y_{J\a} \; 
\oB^{c_1}_{q}
\rt \}
\oB^{c_2}_{p_2}\oB^{c_3}_{p_3}
\ee
In order to be able to construct a dotspinor pseudosuperfield from expression (\ref{wfgreerhjtuyj}) , we need to verify that $d_3$ yields zero when applied to the expression (\ref{wfgreerhjtuyj}):
\be
d_3 
\w_{\dot \a}
=
0=\eb
d_3 
f^{p_1}_{ p_2 p_3}
 \lt \{
\ve^{c_1 c_2 c_3}
g  
\oy_{Q c_1 p_1 i \dot \a} K^i
\; 
B_{c_2}^{p_2} B_{c_3}^{p_3}
+
\ve^{c_1 c_2 c_3}
b_{p_1 q}
\oy_{J\dot \a} \; 
B_{c_1}^{q}
B_{c_2}^{p_2}
B_{c_3}^{p_3}
\rt \} 
\ee
This is not difficult to verify.
We  pick out the terms of $d_3$ that actually have an effect in this case:
\be
d_3 
\w_{\dot \a}
=
f^{p_1}_{ p_2 p_3}
\eb \oC_{\dot \a} 
\lt \{
g \e_{ij} H^i K^j  
\oy_{J \dot \a}^{ \dag}
+
\lt (
t_{p  q} \e_{ij}   K^j T_c^{ q}
+b_{p  q} \e_{ij}   H^j B_c^{ q}
\rt )
\oy_{Q c p i\dot \a}^{ \dag}
\rt \}
\eb
\lt \{
\ve^{c_1 c_2 c_3}
g  
\oy_{Q c_1 p_1 i \dot \a} K^i
\; 
B_{c_2}^{p_2} B_{c_3}^{p_3}
-
\ve^{c_1 c_2 c_3}
b_{p_1 q}
\oy_{J\dot \a} \; 
B_{c_1}^{q}
B_{c_2}^{p_2}
B_{c_3}^{p_3}
\rt \}
\ee
This yields
\be
d_3 
\w_{\dot \a}
=
f^{p_1}_{ p_2 p_3}
 \oC_{\dot \a} 
g \e_{ij} H^i K^j  
\oy_{J \dot \a}^{ \dag}
\eb
\lt \{
+
\ve^{c_1 c_2 c_3}
b_{p_1 q}
\oy_{J\dot \a} \; 
B_{c_1}^{q}
B_{c_2}^{p_2}
B_{c_3}^{p_3}
\rt \}
\eb
+
f^{p_1}_{ p_2 p_3}
 \oC_{\dot \a} 
\lt (
t_{p  q} \e_{ij}   K^j T_c^{ q}
+b_{p  q} \e_{ij}   H^j B_c^{ q}
\rt )
\oy_{Q c p i\dot \a}^{ \dag}
\eb
\lt \{
\ve^{c_1 c_2 c_3}
g  
\oy_{Q c_1 p_1 i \dot \a} K^i
\; 
B_{c_2}^{p_2} B_{c_3}^{p_3}
\rt \}
\ee
which is
\be
d_3 
\w_{\dot \a}
=
f^{p_1}_{ p_2 p_3}
 \oC_{\dot \a} 
g \e_{ij} H^i K^j  
\ve^{c_1 c_2 c_3}
b_{p_1 q}
B_{c_1}^{q}
B_{c_2}^{p_2}
B_{c_3}^{p_3}
\eb
+
f^{p}_{ p_2 p_3}
 \oC_{\dot \a} 
\lt (
t_{p  q} \e_{ij}   K^j T_c^{ q}
+b_{p  q} \e_{ij}   H^j B_c^{ q}
\rt )
\ve^{c_1 c_2 c_3}
g  
 K^i
\; 
B_{c_2}^{p_2} B_{c_3}^{p_3}
\ee
Using the fact that
\be
 \e_{ij} K^i K^j  
=0
\ee
this reduces to
\be
d_3 
\w_{\dot \a}
=
 \oC_{\dot \a} 
g \e_{ij} H^i K^j  
\ve^{c_1 c_2 c_3}
f^{p_1}_{ p_2 p_3} b_{p_1 q}
B_{c_1}^{q}
B_{c_2}^{p_2}
B_{c_3}^{p_3}
\eb
+
 \oC_{\dot \a} 
 \e_{ij} K^i  H^j 
\ve^{c c_2 c_3}
g  
f^{p}_{ p_2 p_3}
b_{p  q}
 \; 
B_c^{ q}
B_{c_2}^{p_2} B_{c_3}^{p_3}
=0
\ee
where we use the antisymmetry of $\e_{ij}$ again.

So we have verified that 
\be
0=
d_3 
\w_{\dot \a}
=
\eb
d_3 
f^{p_1}_{ p_2 p_3}
\ve^{c_1 c_2 c_3}
\lt \{
g  
\oy_{Q c_1 p_1 i \dot \a} K^i
+
b_{p_1 q}
\oy_{J\dot \a} \; 
B_{c_1}^{q}
\rt \}
B_{c_2}^{p_2} B_{c_3}^{p_3}
\ee
and this means that we can construct a dotspinor composite operator in the cohomology space corresponding to this expression, using any of the techniques described in 
\ci{cybersusyII}. We shall discuss this construction below with reference to the simpler example of an electron.

Note that without loss of generality, because of the presence of $\ve^{c_1 c_2 c_3}
$ in the expression above, we can write:
\be 
f^{q}_{ p_2 p_3}
b_{q p_1}=f \ve_{p_1 p_2 p_3}
\ee
and
\be 
f^{q_1}_{ p_2 p_3}
=
f(b^{-1})^{p_1 q_1} \ve_{p_1 p_2 p_3}
\ee
where we define $(b^{-1})^{p_1 q_1}$ so that
\be
(b^{-1})^{p_1 q_1}b_{q_1 s_1}=
\d^{p_1}_{s_1}
\ee

One of the terms in the resulting composite dotspinor expression is the following:
\be 
f(b^{-1})^{p_1 q_1} \ve_{p_1 p_2 p_3}
\ve^{c_1 c_2 c_3}
g  
\oy_{Q c_1 p_1 i \dot \a} K^i
\; 
B_{c_2}^{p_2} B_{c_3}^{p_3}
\Ra
\ee
\be 
f(b^{-1})^{p_1 q_1} \ve_{p_1 p_2 p_3}
\ve^{c_1 c_2 c_3}
g  
\oy_{Q c_1 p_1 i \dot \a} K^i
\; 
\y_{B  c_2}^{p_2} \cdot \y_{B c_3}^{p_3}
\ee
When $K$ takes its VEV:
\be 
K^i \ra v^i m, 
\ee
this yields a  contribution to the $dsb$ quark structure of the $dsb$ $J=\fr{1}{2}$ baryon which is  called the $\X^-_b$ on page 994 and 995 of \ci{physletreview} .  In the limit where $b_{q p_1}$ is diagonal and the CKM angles are all zero, this would be a pure state of the $dsb$ kind.

There are similar expressions for the proton, the  neutron, their generalizations for three flavours, and the $uct$ baryons, all of which give rise to composite dotspinors in the cohomology space.

Two remarks are in order here, for the sake of clarity.  One could have looked for composite dotspinors that are not color or $SU(2)$ singlets, and these probably exist.  But they would not be in the cohomology space of the full BRS operator, which can clearly only contain singlets of these quantum numbers. The same reasoning would exclude $U(1)$ quantum numbers, but we include them anyway. So there has been some tampering with the choice of interesting operators, with an eye towards the physics, and a realization that when gauge symmetry is included, there will be a restriction to color SU(3) and weak isospin  SU(2) singlets. The weak isospin-hypercharge gauge symmetry $SU(2) \times U(1)$ will be broken down to electric charge $U(1)$ by the VEV of course.

\subsection{An example of a electron-type dotspinor from the cohomology space}

\la{qwergregqergergerlep}

Now let us consider a simple generator for a leptonic solution to the constraint equations. 

Consider the following expression which we used in \ci{cybersusyI} (actually we used its complex conjugate):
\be
\ov \w^q_{\a}=
\og 
 \y^{qi}_{L \a}   \oK_i \; 
+
\ov p^{q  p  }
\y_{J\a} \; 
\ov P_{ p }
\la{fwefewfw}
\ee
 Note that it has lepton number L=1, hypercharge Y=-2, and that  it will give rise to  an antichiral undotted spinor pseudosuperfield with spin $J=\fr{1}{2}$, provided the constraint is satisfied.  These are the quantum numbers of the electron.
The relevant part of $d_3 $ here is 
\be
d_3 
= \oC_{\dot \a} 
\lt \{
\og \e^{ij} \oH_i \oK_j  
\y_{J \a}^{ \dag}
\ebp
+
\lt (
\ov p^{p q} \e^{ij}  \oH_j \oP_{ q} 
+ \ov r^{p q} \e^{ij}  \oK_j \oR_{ q}
\rt )
\y_{L  \a}^{p i \dag}
\rt \}
\ee
The demonstration that $d_3$ yields zero is very similar to the demonstration for the hadron in subsection \ref{qwergregqergerger}:
\be
d_3 
\ov \w^q_{ \a}
=
 \oC_{\dot \a} 
\eb
\lt \{
\og \e^{ij} \oH_i \oK_j  
\y_{J \a}^{ \dag}
+
\lt (
\ov p^{p q} \e^{ij}  \oH_j \oP_{ q} 
+ \ov r^{p q} \e^{ij}  \oK_j \oR_{ q}
\rt )
\y_{L  \a}^{p i \dag}
\rt \}
\eb
\lt \{
\og 
  \y^{qi}_{L \a}   \oK_i \; 
+
\ov p^{q  p  }
\y_{J\a} \; 
\ov P_{ p }
\rt \}
\eb
=
 \oC_{\dot \a} 
\lt \{
\og \e^{ij} \oH_i \oK_j  
\y_{J \a}^{ \dag}
\rt \}
\lt \{
\ov p^{q  p  }
\y_{J\a} \; 
\ov P_{ p }
\rt \}
\eb
+
 \oC_{\dot \a} 
\lt \{
\lt (
\ov p^{p q} \e^{ij}  \oH_j \oP_{ q} 
\rt )
\y_{L  \a}^{p i \dag}
\rt \}
\lt \{
\og 
  \y^{qi}_{L \a}   \oK_i \; 
\rt \}
=0
\ee

So we have shown that this can be built up into an antichiral undotted spinor pseudosuperfield, which means that its complex conjugate is a chiral dotted spinor pseudosuperfield, as claimed in \ci{cybersusyI}. We will return below to the specific form of this superfield and its components.

\section{Full Form of one of the Dotspinor Pseudosuperfields for the Electron}

We want to start with the expression 
\be
\ov \w^q_{ E\a}=
\og 
 \y^{qi}_{L \a}   \oK_i \; 
+
\ov p^{q  p  }
\y_{J\a} \; 
\ov P_{ p }
\la{fwefewfw222}
\ee
and build it up into its full components, using the methods outlined in section 9 of \ci{cybersusyII}.

So we need the two fundamental  antichiral composite undotted  spinor pseudosuperfields that arise from the two
spinors in (\ref{fwefewfw222}).  These are:

\be
{\hat {\ov \f}}^{qi}_{L \a}(x) =
 \y^{qi}_{L \a}(\ov y) +
\ov \q^{\dot \b} 
\lt (
\pa_{\a \dot \b} A^{qi}_{L}(\ov y) 
+  C_{\a}  \ov Y^{qi}_{L \dot \b}(\ov y) \rt )
- \fr{1}{2} \ov \q^{\dot \g} \ov \q_{\dot \g} 
\ov \G^{qi}_{L}(\ov y)  
C_{ \a} 
\la{weirdcc}
\ee

\be
{\hat {\ov \f}}^{}_{J \a}(x) =
 \y^{qi}_{J \a}(\ov y) +
\ov \q^{\dot \b} 
\lt (
\pa_{\a \dot \b} A^{}_{J}(\ov y) 
+  C_{\a}  \ov Y^{}_{J \dot \b}(\ov y) \rt )
- \fr{1}{2} \ov \q^{\dot \g} \ov \q_{\dot \g} 
\ov \G^{}_{J}(\ov y)  
C_{ \a} 
\la{weirdcc}
\ee

and the standard notation is:
\be
{\widehat {\ov \f}}^{qi}_{L  \a}(x)=  
\ov \f^{qi}_{L\a}(\ov y) +  \ov \q^{\dot \d} \ov W^{qi}_{L, \d \dot \a}(\ov y)+ 
\fr{1}{2} \ov \q \cdot \ov \q \ov \L^{qi}_{L,  \a}(\ov y)
\ee
So
we identify
\be
{\ov \f}^{qi}_{L \a  }   = 
\y^{qi}_{L \a } 
\la{fund1cc}
 \ee
\be
{\ov W}^{qi}_{L,\a \dot \b}   =
\lt (
\pa_{\a \dot \b} A^{qi}_{L} 
+  C_{\a} \ov Y^{qi}_{ L \b}  \rt )
\la{fund2cc}
\ee
\be
{\ov \L}^{qi}_{L,\a }   =
-\ov \G^{qi}_{L}   
C_{ \a} 
\la{fund3cc}
\ee

Next we need the antichiral scalar pseudosuperfields that correspond to the two scalar fields in (\ref{fwefewfw222}) as discussed in subsection 6.1 of \ci{cybersusyII}.  These are: 
\be
{\hat \oK}_{i}(x) = \oK_{i}(\ov y)
+
\oq^{\dot \b} 
\oy_{K,i \dot \b}(\ov y)
+ \fr{1}{2} \q^{\g} \q_{\g} 
\ov G_{K,i}(\ov y)
\la{qgregerjtjt3}
\ee
and
\be
{\hat \oP}_{p}(x) = \oP_{p}(\ov y)
+
\oq^{\dot \b} 
\oy_{P,p \dot \b}(\ov y)
+ \fr{1}{2} \q^{\g} \q_{\g} 
\ov G_{P,p}(\ov y)
\la{qgregerjtjt3}
\ee

Then the total expression is
\be
{\hat {\ov \w}}^q_{E \a}=
\og 
 {\hat {\ov \f}}^{qi}_{L \a}   {\hat  \oK}_{i }
+
\ov p^{q  p  }
{\hat {\ov \f}}_{J\a} \; 
{\hat {\ov P}}_{ p }
\la{fwefewfw222wwe}
\ee
This is antichiral:
\be
D_{\b} {\hat {\ov \w}}^q_{ \a}=0
\ee
and its components can be found by projection, as was done in \ci{cybersusyII}, using the methods of \ci{superspace}.  The difference is that here the expression actually satisfies the constraint identically, whereas  in \ci{cybersusyII} we needed to assume that it satisfied the constraint, but we had no actual example.  The projection has the form:
\be
{ {\ov \w}}^q_{E \a}=
\lt \{ \og 
 {\hat {\ov \f}}^{qi}_{L \a}   {\hat  \oK}_{i }
+
\ov p^{q  p  }
{\hat {\ov \f}}_{J\a} \; 
{\hat {\ov P}}_{ p } \rt \}_|
\eb
=
\og 
 \y^{qi}_{L \a}   \oK_i \; 
+
\ov p^{q  p  }
\y_{J\a} \; 
\ov P_{ p }\ee
\be
{\ov W}^q_{E\a \dot \b }=
\ov D_{\dot \b }\lt \{ \og 
 {\hat {\ov \f}}^{qi}_{L \a}   {\hat  \oK}_{i }
+
\ov p^{q  p  }
{\hat {\ov \f}}_{J\a} \; 
{\hat {\ov P}}_{ p }\rt \}_|
\ee
\be
{  {\ov \L}}^q_{E \a}=
\fr{1}{2} \ov D^{\dot \b }\ov D_{\dot \b }\lt \{ \og 
 {\hat {\ov \f}}^{qi}_{L \a}   {\hat  \oK}_{i }
+
\ov p^{q  p  }
{\hat {\ov \f}}_{J\a} \; 
{\hat {\ov P}}_{ p }\rt \}_|
\ee
The component forms can be quickly found in this way.
In these four papers on cybersusy, we have little actual need for the component form of any of these expressions, since the actual physics that we find is found using the effective fields.  All we need to know is that these expressions exist, and their quantum numbers and that they satisfy the cybersusy algebra. The latter does require some knowledge of their component forms, as will be seen below.

\subsection{Spontaneous Breaking of Gauge Symmetry}

\la{fqerghjiortbhbrt}

Next we want to derive the algebra used in \ci{cybersusyI} to generate supersymmetry breaking.

The term $- g_{\rm J} m^2 J$ in
Table (\ref{fqwefweef1212}) gives rise to gauge symmetry breaking if $g_{\rm J}\neq 0$. We note that
\be
\ov F'_J = \fr{\pa \; P_{{\rm SP}}  }{\pa J} = 
\ve_{ij} 
g  H^i K^j 
- g_{\rm J} m^2 
\ee
needs a shift of the scalar field parts of the superfield to eliminate the $m^2$ term:
\be
H^1 \ra (mv + H^1)
, 
K^2 \ra (mv + K^2)
\la{substtt}
\ee
Then we have a zero vacuum expectation value for the auxiliary field:
\be
<\ov G_J>_{VEV} = 0  = \ve_{ij} 
g < H^i K^j >_{VEV}
- g_{\rm J} m^2 
\ee
which means that supersymmetry is conserved by this VEV.

This is the development of a vacuum expectation value (VEV) of $m v$ in these two fields, followed by a shift to fields with no vacuum expection value.  Here we have 
\be
v^2 = \fr{g_{\rm J}}{g}
\la{tgqergreojirge}
\ee 

We need to rewrite the action after the substitution 
in equation (\ref{substtt}), in terms of new eigenstates of mass and charge.

For example, in terms of the shifted fields, we define the mass eigenstates:
\be
K = \fr{1}{\sqrt{2}}
\lt ( H^1 - K^2
\rt )
\ee
\be
H = \fr{1}{\sqrt{2}}
\lt ( H^1 + K^2
\rt )
\ee
This has the inverse
\be
K^2 = \fr{1}{\sqrt{2}}\lt (H-K\rt )
\ee
\be
H^1 = \fr{1}{\sqrt{2}}\lt (H+K\rt )
\ee
In terms of the original action, however, this amounts to the substitution:
\be
K^2 \ra m v +  \fr{1}{\sqrt{2}}\lt (H-K\rt )
\ee
\be
H^1 \ra mv +  \fr{1}{\sqrt{2}}\lt (H+K\rt )
\ee

The  superpotential, after this shift and redefinition, written in terms of mass eigenstates is then:
\[
P_{\rm SSP } =
\]
\[
  \lt ( g  m v  \sqrt{2} \; H +   g   \fr{1}{2} ( H H - K K ) -g  H^- K^+ \rt )J
\]
\[
+
p_{pq} \lt [ N^{p} H^- -E^{p} \lt \{ mv + \fr{1}{\sqrt{2}}\lt (H+K\rt )\rt \} \rt ]
 P^{q}
\]
\[
+
r_{pq} \lt [ N^{p} \lt \{ mv + \fr{1}{\sqrt{2}}\lt ( H-K\rt )
\rt \} -E^{p} K^+ \rt ]
 {R}^{q} 
\]
\[
+
t_{pq} \lt [ U^{c p } \lt \{ mv + \fr{1}{\sqrt{2}}\lt (H-K\rt )
\rt \} -D^{c p } K^+ \rt ] 
T_c^q
\]
\be
+
b_{pq} \lt [ U^{c p } H^- -D^{c p } \lt \{ mv + \fr{1}{\sqrt{2}}\lt (H+K\rt )\rt \} \rt ] 
 B_c^q
\la{gqeghrtrtwhrth}
\ee

In the above $H$ and $J$ constitute a massive Higgs boson supermultiplet, and $K,K^{+},H^{-}$ are three Goldstone Boson supermultiplets which will be `eaten' by the vector boson supermultiplets to form the massive weak vector boson supermultiplets $Z^{0}, W^{+},W^{-}$.

The following summarizes our notation for the the supersymmetric standard model (SSM) with Spontaneously Broken Gauge   Symmetry (SBGS).

\be
{\normalsize\begin{tabular}{|c|c|c|c|c|c|c|c|c|}
\hline
\multicolumn{9}{|c|}{\bf  Matter and Zinn Fields after SBGS}
\\
\hline
 Field & Old &  Q& L& B  
& $\y_{\a}$ & $G$ & $\oG$ & $\oY_{\dot \b}$
\\
\hline
$ E^{p}$
 &
$L^{p2}$
&
-1
&
1
&
0
& $\y_{E,\a}^{p}$ & $G_E^{p}$ 
& $\oG_E^{p}$ & $\oY_{E, \dot \b}^{p}$
\\
\hline
$ P^{p}$
 &
$P^{p}$
&
$ 1$
&
$ -1$
&
$ 0$
& $\y_{P,\a}^{p}$ & $G_P^{p}$ 
& $\oG_P^{p}$ & $\oY_{P, \dot \b}^{p}$
\\
\hline
$ R^p $
 &
$R^p $
&
$ 0$
&
$ -1$
&
$ 0$
& $\y_{R,\a}^{p}$ & $G_{R}^{p}$ 
& $\oG_{R}^{p}$ & $\oY_{R, \dot \b}^{p}$
\\
\hline
$ N^{p}$
 &
$L^{p1}$
&
$ 0$
&
$ 1$
&
$ 0$
& $\y_{N,\a}^{p}$ & $G_{N}^{p}$ 
& $\oG_{N}^{p}$ & $\oY_{N, \dot \b}^{p}$
\\
\hline
$ J$
 &
$J$
&
$ 0$
&
$ 0$
&
$ 0$
& $\y_{J,\a}^{p}$ & $G_{J}^{p}$ 
& $\oG_{J}^{p}$ & $\oY_{J, \dot \b}^{p}$
\\
\hline
$ H$
 &
$  \fr{1}{\sqrt{2}}
\lt ( H^1 + K^2
\rt )
$
&
$ 0$
&
$ 0$
&
$ 0$
& $\y_{H,\a}$ & $G_{H}$ 
& $\oG_{H}$ & $\oY_{H, \dot \b}$
\\
\hline
$ H^{-}$
 &
$H^2$
&
$ -1$
&
$ 0$
&
$ 0$
& $\y_{H,\a}^{-}$ & $G_{H}^{-}$ 
& $\oG_{H}^{-}$ & $\oY_{H, \dot \b}^{-}$
\\
\hline
$ K$
 &
$ \fr{1}{\sqrt{2}}
\lt ( H^1 - K^2
\rt )
$
&
$ 0$
&
$ 0$
&
$ 0$
& $\y_{K,\a}$ & $G_{K}$ 
& $\oG_{K}$ & $\oY_{K, \dot \b}$
\\
\hline
$ K^{+}$
 &
$K^1$
&
$ +1$
&
$ 0$
&
$ 0$
& $\y_{K,\a}^{+}$ & $G_K^{+}$ 
& $\oG_K^{+}$ & $\oY_{K, \dot \b}^{+}$
\\
\hline
$ U^{cp}$
 &
$Q^{cp1}
$
&
$ \fr{2}{3}$
&
$ 0$
&
$ \fr{1}{3}$
& $\y_{U,\a}^{cp}$ & $G_{U}^{cp}$ 
& $\oG_{U}^{cp}$ & $\oY_{U, \dot \b}^{cp}$
\\
\hline
$ T^{p}_{c}$
 &
$ T^{p}_{c}$
&
$ -\fr{2}{3}$
&
$ 0$
&
$ -\fr{1}{3}$
& $\y_{T,c,\a}^{p}$ & $G_{T,c}^{p}$ 
& $\oG_{T,c}^{p}$ & $\oY_{T,c, \dot \b}^{p}$
\\
\hline
$ D^{c,p}$
 &
$Q^{cp2}$
&
$ -\fr{1}{3}$
&
$ 0$
&
$ \fr{1}{3}$
& $\y_{D,\a}^{cp}$ & $G_{D}^{cp}$ 
& $\oG_{D}^{cp}$ & $\oY_{D, \dot \b}^{cp}$
\\
\hline
$ B^{p}_{c}$
 &
$ B^{p}_{c}$
&
$ \fr{1}{3}$
&
$ 0$
&
$ -\fr{1}{3}$
& $\y_{B,c,\a}^{p}$ & $G_{B,c}^{p}$ 
& $\oG_{B,c}^{p}$ & $\oY_{B,c, \dot \b}^{p}$
\\
\hline
\end{tabular}
\\
}\ee

\be
\begin{tabular}{|c|c|c|c|c|}
\hline
\multicolumn{5}{|c|}{\bf  CC Matter Fields after SBGS}
\\
\hline
 Field & Old &  Q& L& B  
\\
\hline
$ \oE^{}_{p}$
 &
$\oL_{p2}$
&
1
&
-1
&
0
\\
\hline
$ \oP^{}_{p}$
 &
$\oP^{}_{p}$
&
$ -1$
&
$ 1$
&
$ 0$
\\
\hline
$ \ov R_q $
 &
$\ov R_q $
&
$ 0$
&
$ 1$
&
$ 0$
\\
\hline
$ \oN_{p}$
 &
$\oL_{p1}$
&
$ 0$
&
$ -1$
&
$ 0$
\\
\hline
$ \oJ$
 &
$\oJ$
&
$ 0$
&
$ 0$
&
$ 0$
\\
\hline
$ \oH$
 &
$  \fr{1}{\sqrt{2}}
\lt ( \oH_1 + \oK_2
\rt )
$
&
$ 0$
&
$ 0$
&
$ 0$
\\
\hline
$ \oH^{-}$
 &
$\oH_2$
&
$ 1$
&
$ 0$
&
$ 0$
\\
\hline
$ \oK$
 &
$ \fr{1}{\sqrt{2}}
\lt ( \oH_1 - \oK_2
\rt )
$
&
$ 0$
&
$ 0$
&
$ 0$
\\
\hline
$ \oK^{-}$
 &
$\oK_1$
&
$ -1$
&
$ 0$
&
$ 0$
\\
\hline
$ \oU_{cp}$
 &
$\oQ_{cp1}
$
&
$ -\fr{2}{3}$
&
$ 0$
&
$ -\fr{1}{3}$
\\
\hline
$ \oT_{p}^{c}$
 &
$ \oT_{p}^{c}$
&
$ +\fr{2}{3}$
&
$ 0$
&
$ +\fr{1}{3}$
\\
\hline
$ \ovD_{c,p}$
 &
$\oQ_{cp2}$
&
$ +\fr{1}{3}$
&
$ 0$
&
$ -\fr{1}{3}$
\\
\hline
$ \oB_{p}^{c}$
 &
$ \oB_{p}^{c}$
&
$ -\fr{1}{3}$
&
$ 0$
&
$ \fr{1}{3}$
\\
\hline
\end{tabular}
\ee

\section{Cohomology for Leptons   in the Standard Model, and the Introduction of Spontaneous Breaking}

\la{lepappendix}

\subsection{Discussion}

Up to now we have discussed the BRS cohomology in general for a massless chiral scalar action, and then we have exhibited, in subsections 
\ref{qwergregqergerger} and \ref{qwergregqergergerlep}, two solutions of the constraint equations in a specific example-the massless SSM. For the hadronic case we looked at expression (\ref{wfgreerhjtuyj}), which  was
\be
\w_{\dot \a}
=
f^{p_1}_{ p_2 p_3}
\ve^{c_1 c_2 c_3}
\lt \{
g  
\oy_{Q c_1 p_1 i \dot \a} K^i
+
b_{p_1 q}
\oy_{J\dot \a} \; 
B_{c_1}^{q}
\rt \}
B_{c_2}^{p_2} B_{c_3}^{p_3}
\ee
and we noted that it could be written in the form
\be
\w_{\dot \a}
=
f \ve^{c_1 c_2 c_3}
\lt \{
   (b^{-1})^{p_1 q_1} \ve_{q_1 p_2 p_3}
g \oy_{Q c_1 p_1 i \dot \a} K^i
+
\ve_{p_1 p_2 p_3}
\oy_{J\dot \a} \; 
B_{c_1}^{q}
\rt \}
B_{c_2}^{p_2} B_{c_3}^{p_3}
\ee
It satisfies
\be
d_3 {\w}_{\dot \a} = 0
\la{fqewfwe1}
\ee
For the leptonic case we looked at
expression (\ref{qwergregqergergerlep}), which was:
\be
\ov \w^q_{ \a}=
\og 
 \y^{qi}_{L \a}   \oK_i \; 
+
\ov p^{q  p  }
\y_{J\a} \; 
\ov P_{ p }
\la{qwergregqergergerlep1}\ee
It satisfies:
\be
d_3 {\ov \w}_{\a} = 0
\la{fqewfwe2}
\ee

What happens to these solutions of the constraint equations when the  Spontaneous Breaking of gauge symmetry in the SSM is turned on?  

The answer is that the right hand side of the equations 
(\ref{fqewfwe1}) and (\ref{fqewfwe2}) for these hadronic and leptonic cases is no longer zero after gauge symmetry breaking.  It is in fact something more interesting than zero. 

Explicitly, in the spontaneously broken theory, we find that $d_3$ is equation 
(\ref{egrggwrtrhtrhbntr}) except that the coefficient of $\oy_{J \dot \a}^{ \dag}$ is changed:
\be
 \oC_{\dot \a} 
 \lt (
g \e_{ij} H^i K^j  \rt )
\oy_{J \dot \a}^{ \dag}
\Ra
 \oC_{\dot \a} 
 \lt ( -g_J m^2 +
g \e_{ij} H^i K^j  \rt )
\oy_{J \dot \a}^{ \dag}
\ee
Since we are using this in the context where there are no derivatives, we can actually use the above and change variables later.  But it is a comfort to use the new variables too, to be sure that things are working correctly.

Let us focus on the electron type operator in equation  (\ref{qwergregqergergerlep1}).   

\subsection{Calculation with $d_3$ in the   theory with SBGS  }

Let us return to the treatment of the operator in equation  (\ref{qwergregqergergerlep1}) that was done in section \ref{qwergregqergergerlep}, but now let us perform the shift and change of variables that are appropriate to the case with spontaneously broken gauge symmetry.  The notation and procedure for the shift and change of variables here are explained in section \ref{fqerghjiortbhbrt}.

\be
\ov \w^q_{P \a}=
\og 
\y^{q}_{L \a} \cd \oK \; 
-
\ov p^{q  p  }
\y_{J\a} \; 
\ov P_{ p }
\Ra
\ee
\[
\ov \w^q_{P \a}=
-\og 
\y^{-q}_{E \a} \cd
 \lt \{
m \ovv + 
\fr{1}{\sqrt{2}}
\lt (
\oH - {\oK}
\rt )
\rt \}
+
\y^{q}_{N \a} \cd  
\oK^- \; 
\]\be
-
\ov p^{q  p  }
\y_{J\a} \; 
\ov P_{ p }
\la{ghwruuyjkdf}
\ee

After SBGS, the operator $d_3$, and the full pseudosuperfields that it corresponds to, are of the form

\[
d_3 
= \oC_{\dot \a} 
\lt \{
\fr{\pa \; P_{{\rm SSP}} }{\pa K^i } 
\oy_{K i\dot \a}^{ \dag}
\rt.
\]
\[
+
\fr{\pa \; P_{{\rm SSP}} }{\pa H^i } 
\oy_{H i \dot \a}^{ \dag}
+
\fr{\pa \; P_{{\rm SSP}} }{\pa J } 
\oy_{J\dot \a}^{ \dag}
\]
\[
+
\fr{\pa \; P_{{\rm SSP}} }{\pa L^{p i} } 
\oy_{L i \dot \a}^{p  \dag}
+
\fr{\pa \; P_{{\rm SSP}} }{\pa   P^{ q}  } 
\oy_{P\dot \a}^{ \dag}
\]
\[
+
\fr{\pa \; P_{{\rm SSP}} }{\pa   R^{ q}  } 
\oy_{R q\dot \a}^{q \dag}
+
\fr{\pa \; P_{{\rm SSP}} }{\pa Q^{c p i} } 
\oy_{Q c p i\dot \a}^{ \dag}
\]
\[
\lt.
+
\fr{\pa \; P_{{\rm SSP}} }{\pa    T_c^{ q}  } 
\oy_{T  q\dot \a}^{c  \dag}
+
\fr{\pa \; P_{{\rm SSP}} }{\pa   B_c^{ q}  } 
\oy_{B q\dot \a}^{c \dag}
\rt \}
\]
\be
+ {\rm Complex\; Conjugate}
\la{fwerwefgwfgwe}
\ee
where 
$P_{{\rm SSP}}$ is defined in equation (\ref{gqeghrtrtwhrth}).

Also we must reexpress the derivatives and the superpotential in terms of the shifted mass eigenstate fields.

Equation (\ref{fwerwefgwfgwe})
 takes the place of equation (\ref{qefwefwefwefopjkpjopj}), and it summarizes what happens to the relevant full fundamental dotspinor pseudosuperfields (which are not actually superfields, as emphasized in \ci{cybersusyII}--they have the quadratic inhomogeneous terms for the theory without SBGS, and when gauge symmetry is broken, they acquire additional terms as discussed below).

For the particular case (\ref{ghwruuyjkdf}), 
we only need the following parts of $d_3$, now with shifted and redefined fields:
 \be
d_{3 {\rm Left \;Lepton}} = C_{\a} \lt \{
 \fr{\pa \; \ov P_{{\rm SSP}} }{\pa \ov E_q } 
  \y_{E \a}^{q \dag}
+
 \fr{\pa \; \ov P_{{\rm SSP}} }{\pa \ov N_q } 
    \y_{N \a}^{q \dag}
+
 \fr{\pa \; \ov P_{{\rm SSP}} }{\pa \ov J } 
 \y_{J \a}^{ \dag}
\rt \}
\ee
where
\be
\fr{\pa \; \ov P_{{\rm SSP}} }{\pa \ov E_q } 
=
-\ov r^{pq}  \ov R_{q} \oK^- 
- \ov p^{pt} \oP_{t}
\lt \{
m \ovv + 
\fr{1}{\sqrt{2}}
\lt (
\oH + {\oK}
\rt )
\rt \}
\ee

\be
\fr{\pa \; \ov P_{{\rm SSP}} }{\pa \ov N_q }  =
+\ov p^{pt}  \oP^-_{t} \oH^+ 
+ \ov r^{pt} \ov R_{t}
\lt \{
m \ovv + 
\fr{1}{\sqrt{2}}
\lt (
\oH - {\oK}
\rt )
\rt \}
\ee

\be
\fr{\pa \; \ov P_{{\rm SSP}} }{\pa \ov J } 
=
\og m \ovv 
\sqrt{2} \; \oH
+
\og  \fr{1}{2}
\oH \oH 
- 
\og  \fr{1}{2}
{\oK} {\oK}
- 
\og 
\oH^+
 \oK^- 
\ee

  In terms of the variables of the spontaneously broken theory, we have:

\be
d_{3 {\rm Left \;Lepton}}=\eb
 C_{\a} \lt [-\ov r^{pq}  \ov R_{q} \oK^- 
- \ov p^{pt} \oP_{t}
\lt \{
m \ovv + 
\fr{1}{\sqrt{2}}
\lt (
\oH + {\oK}
\rt )
\rt \} \rt ]\y_{E \a}^{q \dag}
\eb
-
 C_{\a} \lt [ \ov p^{pt}  \oP^-_{t} \oH^+ 
+ \ov r^{pt} \ov R_{t}
\lt \{
m \ovv + 
\fr{1}{\sqrt{2}}
\lt (
\oH - {\oK}
\rt )
\rt \} \rt ] \y_{N \a}^{q \dag}
\eb
+
 C_{\a} \lt [ \og m \ovv 
\sqrt{2} \; \oH
+
\og  \fr{1}{2}
\oH \oH 
- 
\og  \fr{1}{2}
{\oK} {\oK}
- 
\og 
\oH^+
 \oK^-  \rt ]
 \y_{J \a}^{ \dag}
\ee
Now what does $d_{3 {\rm Left \;Lepton}}$ actually do?  It summarizes the requirement that arises when the supersymmetry operator $\d_{\rm BRS}$ acts on a composite operator, in order for that composite operator to behave like a dotspinor.

To see if this expression behaves like a dotspinor, we examine:
\be
d_3 \ov \w^q_{P \a}=
d_3 \lt \{
-\og 
\y^{-q}_{E \a} \cd
 \lt \{
m \ovv + 
\fr{1}{\sqrt{2}}
\lt (
\oH - {\oK}
\rt )
\rt \}
+
\y^{q}_{N \a} \cd  
\oK^- \; 
-
\ov p^{q  p  }
\y_{J\a} \; 
\ov P_{ p }
\rt \}
\ee
and this yields
\be
d_3 \ov \w^q_{P \a}=
-\og 
C_{\a} \lt [-\ov r^{pq}  \ov R_{q} \oK^- 
- \ov p^{pt} \oP_{t}
\lt \{
m \ovv + 
\fr{1}{\sqrt{2}}
\lt (
\oH + {\oK}
\rt )
\rt \} \rt ] \cd  
\eb
\lt \{
m \ovv + 
\fr{1}{\sqrt{2}}
\lt (
\oH - {\oK}
\rt )
\rt \} 
\eb
-
 C_{\a} \lt [ \ov p^{pt}  \oP^-_{t} \oH^+ 
+ \ov r^{pt} \ov R_{t}
\lt \{
m \ovv + 
\fr{1}{\sqrt{2}}
\lt (
\oH - {\oK}
\rt )
\rt \} \rt ] \cd  \oK^- \; 
\eb
-
\ov p^{q  p  }
 C_{\a} \lt [ \og m \ovv 
\sqrt{2} \; \oH
+
\og  \fr{1}{2}
\oH \oH 
- 
\og  \fr{1}{2}
{\oK} {\oK}
- 
\og 
\oH^+
 \oK^-  \rt ]
\; 
\ov P_{ p }
\ee
and this is summarized by:
\be
d_3 \ov \w^q_{P \a}=
-\og 
C_{\a} \lt [
- \ov p^{pt} \oP_{t}
m \ovv   \rt ] \cd  \lt \{
m \ovv  
\rt \} 
\eb
=
\og m^2 \ovv^2
C_{\a} 
 \ov p^{pt} \oP_{t}  
\la{fwfwefwef}
\ee

It does not behave like a dotspinor.  It is not a superfield.  It has a homogeneous term added to the usual transformation of a chiral dotted spinor superfield.  This is the cybersusy algebra for this operator.

What happens to this operator in equation (\ref{fwfwefwef})
 is summarized in the top line of Table (\ref{qrtewfwuiprtreheh99}):
The other three lines in  Table (\ref{qrtewfwuiprtreheh99}) are similar operators that are in the cohomology space before gauge symmetry breaking and that have the variations shown after GSB.

\vspace{.2cm}
\be
\begin{tabular}{|c|c|c|c|c|c|}
\hline
\multicolumn{6}{|c|}{ Table 
\ref{qrtewfwuiprtreheh99}}
\\
\hline
\multicolumn{6}{|c|}{ Electrons: Generating Expressions for the dotspinor    
}
\\
\multicolumn{6}{|c|}{ multiplets  and their variations when gauge symmetry is broken}
\\
\hline
Spinor
&L
&{\rm S} 
& Ghost=0
& Ghost=1
& Scalar
\\
\hline
$\ov \w^q_{P \a}$
&
$+1$ 
&
$\fr{1}{2}$
&
\bt
$
\og 
\y^{q}_{L \a} \cd \oK \; 
$\\$
-
\ov p^{q  p  }
\y_{J\a} \; 
\ov P_{ p }
$
\et
&
\bt
$
\og m^2 \ovv^2
C_{\a} 
$\\$
 \; 
\ov p^{ q  p }
\ov P_{ p }
$
\et
&
$\ov P_{ p }$
\\
\hline
$\w_{P q \dot \a}$
&$-1$ 
&$\fr{1}{2}$
&
\bt
$
g 
\oy_{L q  \dot \a} \cd K \; 
$\\$
-
p_{ q  p }\oy_{J\dot \a} \; 
P^{ p }
$
\et
&
\bt
$
p_{ q  p }
g v^2 
$\\$
m^2 \oC_{\dot \a} \; 
P^{ p }
$
\et
&$P^{ p }$
\\
\hline
$\ov \w^{ p }_{E\a}$
&$-1$
&$\fr{1}{2}$
&
\bt
$
\og 
\y^{ p }_{P \a} \oK \cd \oH \; 
$\\$
-
(p^{\dag})^{ p  q}
\y_{J\a} \; 
\ov L^{}_{q} \cd \oH
$
\et
&
\bt
$
(p^{\dag})^{ p  q}
\og \ovv^2 
$\\$
m^2 C_{\a} \; 
\ov L^{}_{q} \cd \oH
$
\et
&
$\ov E_{q}$\\
\hline
$\w_{E  p \dot \a}$
&$+1$
&$\fr{1}{2}$
&
\bt
$
g 
\oy^{}_{P  p  \dot \a} K \cd H \; 
$\\$
-
p^T_{   p  q}
\oy_{J\dot \a} \; 
L^{q}  \cd H
$
\et
&
\bt
$
p^T_{   p  q}
g v^2 
$\\$
m^2 \oC_{\dot \a} \; L^{q}  \cd H
$
\et
&
$E^{q}$\\

\hline
\hline

\end{tabular}
\\
\la{qrtewfwuiprtreheh99}
\ee

\vspace{.2cm}

The point is that these expressions behave like dotspinors except for the extra term.  We used these extra terms to generate the new algebra, and the cybersusy action, and the supersymmetry breaking in \ci{cybersusyII}.

\subsection{Neutrinos}

Similarly we have Table (\ref{qrtewfwuipree}).  It lists  four operators with the quantum numbers of the neutrino and its antiparticle which are in the cohomology space before gauge symmetry breaking and that have the variations shown after GSB.

\vspace{.2cm}
\be
\begin{tabular}{|c|c|c|c|c|c|c|}
\hline
\multicolumn{7}{|c|}{ Table 
\ref{qrtewfwuipree}
}
\\
\hline
\multicolumn{7}{|c|}{ Neutrinos: Generating Expressions for the dotspinor    }
\\
\multicolumn{7}{|c|}{ multiplets  and their variations when gauge symmetry is broken}
\\
\hline
Spinor
&L
&{\rm S} 
& Ghost=0
& Ghost=1
& Scalar
& Eff. $\d_{GSB}$
\\
\hline
$\ov \w^q_{R \a}$
&
$+1$ 
&
$\fr{1}{2}$
&
\bt
$
\og 
\y^{q}_{L \a} \cd \oH \; 
$\\$
-
\ov r^{q  p  }
\y_{J\a} \; 
\ov R_{ p }
$
\et
&
\bt
$
\ov r^{ q  p }
\og
\ovv^2
$\\$
 m^2 C_{\a} \; 
\ov R_{ p }
$
\et
&
$\ov R_{ p }$
&
\bt
$  \ov \w^p_{R \a}\ra $\\$
\ov r^{pq}   C_{ \a} \A_{ R q}$
\et
\\
\hline
$\w_{R q \dot \a}$
&$-1$ 
&$\fr{1}{2}$
&
\bt
$
g 
\oy_{L q  \dot \a} \cd H \; 
$\\$
-
r_{ q  p }\oy_{J\dot \a} \; 
R^{ p }
$
\et
&
\bt
$
r_{ q  p }
g v^2 
$\\$
m^2 \oC_{\dot \a} \; 
R^{ p }
$
\et
&$R^{ p }$
&
\bt
$   \w_{R p \dot \a}\ra $\\$
 r_{pq}   \ov C_{\dot \a} A^q_{ R}$
\et\\
\hline
$\ov \w^{ p }_{N\a}$
&$-1$
&$\fr{1}{2}$
&
\bt
$
\og 
\y^{ p }_{R \a} \oK \cd \oH \; 
$\\$
-
(r^{\dag})^{ p  q}
\y_{J\a} \; 
\ov L^{}_{q} \cd \oK
$
\et
&
\bt
$
(r^{\dag})^{ p  q}
\og \ovv^2 
$\\$
m^2 C_{\a} \; 
\ov L^{}_{q} \cd \oK
$
\et
&
$\ov N_{q}$
&
\bt
$  \ov \w^p_{L \a}\ra
$\\$
 \ov r^{pq}   C_{ \a} \A_{ L q}$
\et
\\
\hline
$\w_{N  p \a}$
&$+1$
&$\fr{1}{2}$
&
\bt
$
g 
\oy^{}_{R  p  \dot \a} K \cd H \; 
$\\$
-
r^T_{   p  q}
\oy_{J\dot \a} \; 
L^{q}  \cd K
$
\et
&
\bt
$
r^T_{   p  q}
g v^2 
$\\$
m^2 \oC_{\dot \a} \; L^{q}  \cd K
$
\et
&
$N^{q}$&
\bt
$   \w_{L p \dot \a}\ra 
$\\$
  r_{pq}   \ov C_{ \dot \a} A^q_{ L}$
\et
\\

\hline
\hline

\end{tabular}
\\
\la{qrtewfwuipree}
\ee

\vspace{.2cm}

This is precisely the algebra that we  used in \ci{cybersusyI}.
The same algebra arises for the neutrinos of course, except that the matrices $ r_{ q  p}$ and $ p_{ q   p}$ get exchanged.

\subsection{The observables of the SSM are actually composite Operators}
\la{qfggretthrthrtyhbdb}

Consider the following  terms on the last line in Table 
(\ref{qrtewfwuiprtreheh99}):

\be
\w_{E  p \dot \a} \approx
\lt (
g 
\oy^{}_{P  p  \dot \a} K \cd H \; 
-
p^T_{   p  q}
\oy_{J\dot \a} \; 
L^{q}  \cd H
\rt )
\la{frerghhrthe1}
\eb
\ra 
p^T_{   p  q}
g v^2 
m^2 \oC_{\dot \a} \; L^{q}  \cd H
\ra E^{q}
\la{frerghhrthe2}
\ee
When written in terms of the variables of the SSM with SBGS, this becomes:
\be
 L^{q}  \cd H \ra   N^q H^- - E^{q-} \lt \{ mv + \fr{1}{\sqrt{2}}\lt (H+K\rt ) \rt \}
\la{grertjuyjhrt}.
\ee
Note that there is an elementary field (the Selectron) times a mass, namely $E^{q-}  mv $,  in this composite operator (\ref{grertjuyjhrt}), 
 after SBGS.  The algebra is telling us that this elementary field is really part of a scalar composite operator and that scalar composite operator gets mixed with a dotspinor composite operator after SBGS, and that supersymmetry is broken by that mixing in the way described in \ci{cybersusyI}. 

So the argument would be something like this:
If one ignores the fact that the Selectron  $E^{q-} $ is really a part of a composite operator  (\ref{grertjuyjhrt}), in terms of the variables appropriate after gauge symmetry breaking, and one also ignores the mixing of that operator with the corresponding dotspinor
 $\w_{E  p \a}$ constructed from 
(\ref{frerghhrthe1}), then one misses the supersymmetry breaking.  Of course, this operator $E^{q-} $ is just the scalar of a chiral scalar multiplet, and the above discussion really needs amplification to include the fact that this is a mapping of supermultiplets.

Another point that is very important is that this algebra is not restricted to this simple operator.  The algebra probably needs to be generally true for any operator with the same quantum numbers as the simplest dotspinor for a given quantum number.  Then the argument will work properly.

There is good reason to think that  all relevant operators participate in the cybersusy algebra, as we shall demonstrate in section (\ref{fgqergegeghqweg}) below.

\section{The Baryons with Y=-2 and Y=+2} 

\la{ewtgegeiobhrt}

We now list the generating polynomials for some hadrons in  the spontaneously broken SSM in our notation.  These can be used to construct the composite dotspinors, as was done above for the leptons. 

\be
\begin{tabular}{|c|c|c|c|}
\hline
\multicolumn{4}{|c|}{  Table (\ref{qfqwefqwfqefw}): 
Tower of Cohomology of the Y=-2
}
\\
\multicolumn{4}{|c|}{ baryons when the symmetry is spontaneously broken
}
\\
\hline
{\rm S} 
& D
& Ghost=0
& Ghost=1
\\
\hline
$\fr{1}{2}$
&
$4 \fr{1}{2}$
&
\bt
$
\ve_{c_1 c_2 c_3}
\og f_{p_1}^{ p_2 p_3}
\y^{c_1 p_1}_{Q\a} \cd \oK \; 
\oB^{c_2}_{p_2}\oB^{c_3}_{p_3}
$\\$
-
\ve_{c_1 c_2 c_3}
\ob^{p_1 q}
f_{q}^{ p_2 p_3}
\y_{J\a} \; 
\oB^{c_1}_{p_1}
\oB^{c_2}_{p_2}
\oB^{c_3}_{p_3}
$
\et
&
\bt
$
\ve_{c_1 c_2 c_3}
\ob^{p_1 q}
 f_{q}^{ p_2 p_3}
\og \ovv^2 m^2 C_{\a} \; 
$\\$
\oB^{c_1}_{p_1}
\oB^{c_2}_{p_2}
\oB^{c_3}_{p_3}
$
\et
\\
\hline
$1$
&
$6$
&
\bt
$\ve_{c_1 c_2 c_3}
\og f_{p_1 p_2}^{ p_3}
$\\$
\y^{c_1 p_1}_{Q\a} \cd \oK \; 
\y^{c_2 p_2}_{Q\b} \cd \oK \; 
\oB^{c_3}_{p_3}
$\\$
-
\ve_{c_1 c_2 c_3}
 f_{p_1 q}^{ p_3}
\ob^{p_2 q}
$\\$
\y_{J\a} \y^{c_1 p_1}_{Q\b} \cd \oK \; 
\oB^{c_2}_{p_2}
\oB^{c_3}_{p_3}
$
\et
&
\bt
$
\ve_{c_1 c_2 c_3}
 f_{p_1 q}^{ p_3}
\ob^{p_2 q}m^2 C_{\a}
$\\$
 \y^{c_1 p_1}_{Q\b} \cd \oK \; 
\oB^{c_2}_{p_2}
\oB^{c_3}_{p_3}
$
\et
\\
\hline
$\fr{3}{2}$
&
$7 \fr{1}{2}$
&
\bt
$\ve_{c_1 c_2 c_3}
\og f_{p_1 p_2  p_3}
$\\$
\y^{c_1 p_1}_{Q\a} \cd \oK \; 
\y^{c_2 p_2}_{Q\b} \cd \oK \; 
\y^{c_3 p_3}_{Q\g} \cd \oK \; 
$\\$
-
\ve_{c_1 c_2 c_3}
 f_{p_1   p_2 q}
\ob^{p_3 q}
$\\$
\y_{J\a} \y^{c_1 p_1}_{Q\b} \cd \oK \; 
\y^{c_2 p_2}_{Q\g} \cd \oK \; 
\oB^{c_3}_{p_3}
$
\et
&
\bt
$\ve_{c_1 c_2 c_3}
 f_{p_1   p_2 q}
\ob^{p_3 q}
m^2 C_{\g}
\; 
$\\$
\y^{c_1 p_1}_{Q\a} \cd \oK \; 
\y^{c_2 p_2}_{Q\b} \cd \oK \; 
\oB^{c_3}_{p_3}
$\et
\\
\hline

\end{tabular}
\\
\la{qfqwefqwfqefw}
\ee

There are two recognizable baryons buried in these operators. 
When $K$ takes its VEV, 
\be 
K^i \ra v^i m, 
\ee
the $S = \fr{1}{2}$ operator in Table \ref{qfqwefqwfqefw}
 yields a  contribution to the (antisymmetric) $dsb$ quark structure of the $dsb$ $J=\fr{1}{2}$ baryon which is  called the $\X^-_b$ on page 994 and 995 of \ci{physletreview} .  In the limit where $b_{q p_1}$ is diagonal and the CKM angles are all zero, this would be a pure state of the $dsb$ kind.

The $S = \fr{3}{2}$ operator in Table \ref{qfqwefqwfqefw}
 yields a  contribution to the decuplet of (symmetric)  $ddd,dds,ddb,dss,dsb,dbb, sss,ssb,sbb, bbb$ quark structures of the   $J=\fr{3}{2}$ baryons which are   called the various $\D^-_{J=  \fr{3}{2}} $, depending on the main quarks contained in them. 

The same structure reappears in the following table:

\vspace{.2cm}

\be
\begin{tabular}{|c|c|c|c|}
\hline
\multicolumn{4}{|c|}{ Table (\ref{qtrweiojhpgruig}) : 
Tower of Cohomology of the Y=+2
}
\\
\multicolumn{4}{|c|}{  right antibaryons when the symmetry is spontaneously broken
}
\\
\hline
{\rm Spin} 
& Dim 
& Ghost=0
& Ghost=1
\\
\hline
$\fr{1}{2}$
&
$7 \fr{1}{2}$
&
\bt
$
\ve^{c_1 c_2 c_3}
\og f_{p_1}^{ p_2 p_3}
$\\$
\y^{ p_1}_{B c_1 \a} \; 
\oH \cd \oQ_{c_2  p_2} 
\oH \cd \oQ_{c_3  p_3} 
 \oK \cd \oH
$\\$
-
\ve^{c_1 c_2 c_3}
 f_{q}^{ p_2 p_3}
\ob^{q p_1} 
$\\$
\y_{J \a} \; 
\oH \cd \oQ_{c_1  p_1} 
\oH \cd \oQ_{c_2  p_2} 
\oH \cd \oQ_{c_3  p_3} 
$\et
&
\bt
$
\ve^{c_1 c_2 c_3}
 f_{q}^{ p_2 p_3}
\ob^{q p_1} m^2 C_{\a}
$\\$
 \; 
\oH \cd \oQ_{c_1  p_1} 
\oH \cd \oQ_{c_2  p_2} 
\oH \cd \oQ_{c_3  p_3} 
$
\et
\\
\hline
$1$
&
$7 $
&
\bt
$
\ve^{c_1 c_2 c_3}
\og f_{p_1 p_2}^{ p_3}
$\\$
\y^{ p_1}_{B c_1 \a} \; 
\y^{ p_2}_{B c_2 \b} \; 
\oH \cd \oQ_{c_3  p_3} 
 \oK \cd \oH
$\\$
-
\ve^{c_1 c_2 c_3}
 f_{p_1 q }^{ p_3}
\ob^{q p_2} 
$\\$
\y_{J \a} \; 
\y^{ p_1}_{B c_1 \b} \; 
\oH \cd \oQ_{c_2  p_2} 
\oH \cd \oQ_{c_3  p_3} 
$\et
&
\bt
$
\ve^{c_1 c_2 c_3}
 f_{p_1 q }^{ p_3}
\ob^{q p_2} m^2
C_{ \a} \; 
$\\$
\y^{ p_1}_{B c_1 \b} \; 
\oH \cd \oQ_{c_2  p_2} 
\oH \cd \oQ_{c_3  p_3} 
$
\et
\\
\hline
$\fr{3}{2}$
&
$6 \fr{1}{2}$
&
\bt
$
\ve^{c_1 c_2 c_3}
\og f_{p_1 p_2  p_3}
$\\$
\y^{ p_1}_{B c_1 \a} \; 
\y^{ p_2}_{B c_2 \b} \; 
\y^{ p_3}_{B c_3 \g} \; 
 \oK \cd \oH
$\\$
-
\ve^{c_1 c_2 c_3}
 f_{p_1 p_2   q}
\ob^{q p_3} 
$\\$
\y_{J \a} \; 
\y^{ p_1}_{B c_1 \b} \; 
\y^{ p_2}_{B c_2 \g} \; 
\oH \cd \oQ_{c_3  p_3} 
$\et
&
\bt
$
\ve^{c_1 c_2 c_3}
 f_{p_1 p_2   q}
\ob^{q p_3} 
m^2 C_{ \a} \; 
$\\$
\y^{ p_1}_{B c_1 \b} \; 
\y^{ p_2}_{B c_2 \g} \; 
\oH \cd \oQ_{c_3  p_3} 
$
\et
\\
\hline
\end{tabular}
\\
\la{qtrweiojhpgruig}
\ee

\subsection{The composite baryon operators }

Consider the operator at the left hand top of Table (\ref{qtrweiojhpgruig}):

\be
{\hat \f}^{Y=2,p_1}_{{\cal R}, {\rm Dim}= 7 \fr{1}{2} , p_2 p_3 \a}
\eb
=
\ve^{c_1 c_2 c_3}
\og f_{p_1}^{ p_2 p_3}
\y^{ p_1}_{B c_1 \a} \; 
\oH \cd \oQ_{c_2  p_2} 
\oH \cd \oQ_{c_3  p_3} 
 \oK \cd \oH
\eb
-
\ve^{c_1 c_2 c_3}
 f_{q}^{ p_2 p_3}
\ob^{q p_1} 
\y_{J \a} \; 
\oH \cd \oQ_{c_1  p_1} 
\oH \cd \oQ_{c_2  p_2} 
\oH \cd \oQ_{c_3  p_3} 
\ee
To lowest order in the number of fields, and hence to highest order in the number of factors of the mass m, this operator becomes, after spontaneous breaking of $SU(2) \times U(1)$ down to $U(1)$, the following:

\be
{\hat \f}^{Y=2,p_1}_{{\cal R}, {\rm Dim}= 7 \fr{1}{2} , p_2 p_3 \a}
\eb
\approx \ve^{c_1 c_2 c_3}
\og
\y^{ p_1}_{B c_1 \a} \; 
m \ovD_{c_2  p_2} 
m \ovD_{c_3  p_3} 
 m^2 
\ee
\be
=
m^4 \ve^{c_1 c_2 c_3}
\og
\y^{ p_1}_{B c_1 \a} \; 
 \ovD_{c_2  p_2} 
 \ovD_{c_3  p_3} 
\la{32rewffwef}
\ee
There are also terms like
\be
m^3 \ve^{c_1 c_2 c_3}
\og
\y^{ p_1}_{B c_1 \a} \; 
 \oy_{D c_2  p_2} 
 \cdot \oy_{D c_3  p_3} 
\la{32rewffwef1}
\ee

This is the sort of thing that one would expect for the hadronic operator for the relevant bound state.  When one looks at the rest of the operators, one finds a complicated  system of composite operators with a complicated algebra relating them.  Similar algebras and towers of cohomology exist for the quantum numbers of the proton and neutron and $\S^{++}$. These algebras are similar to the algebra set out above for the leptons, and we hope to return to them and the related effective action for supersymmetry breaking  in a future paper. At the present time, of course, it is not at all clear that the result for the baryons will be phenomenologically viable--the surprise so far is that the corresponding case for the leptons looks quite promising, as seen in \ci{cybersusyI}.

\section{A multiplicity of operators for a given composite particle}
\la{fgqergegeghqweg}

One possible, and reasonable,  objection is that we have abstracted from one operator for a composite electron, and its algebra upon gauge symmetry breaking, to build an effective theory of supersymmetry breaking.  This may look like an error. 

Moreover it is proposed to do the same kind of exercize for the baryons.

It is certainly true that any composite particle, such as a baryon, has certainly got a multitude of operators that can be used to create it, formed from the constituent operators  in the basic theory in which it appears to be a composite.

We shall assume here that cybersusy will give reasonable results for the baryons, in the same way that it has done for the leptons.  If that does not happen, it will be a major problem for cybersusy. Let us suppose that the results are reasonable. 

In order to justify cybersusy as a reasonable supersymmetry breaking mechanism, do we need to establish that all possible operators that can create a given composite, say a proton supermultiplet, in the spontaneously broken SSM, give rise to the same algebra in some sense?  This seems correct, because if there are some such operators that do not give rise to the cybersusy algebra, but do create protons, then how can we know whether to use the effective action or not?

Actually it does seem possible that this idea is right. It is quite possible that all composite operators in the massless SSM which have the right quantum numbers to  create protons do in fact have the cybersusy algebra upon SBGB.

First of all, let us restrict our attention to operators that create the proton supermultiplet, and that are also in the cohomology space, and are also generated from simple generators. 

Even better, let us look at the baryonic tower discussed above in section , rather than the more complicated case of the proton. That tower of  operators descend from an operator that, before supersymmetry breaking,  creates a supermultiplet including the precursor to the $\D^{-}$, with spin $J=\fr{3}{2}$.

 So let us look at this  $\D^{-}$ type 
pseudosuperfield, and call it 
${\hat {\ov \w}}^{-}_{\D,\a \b \g}$.

Now consider the pseudosuperfield 
\be
{\hat {\ov \w}}^{-}_{\D \; {\rm Comp},\a \b \g}= 
{\hat {\ov \w}}^{-}_{\D,\a \b \g} {\hat {\ov A}}_{\rm Composite}
\ee
where ${\hat {\ov A}}_{\rm Composite}$ is any composite scalar pseudosuperfield with any dimension, so long as it has the same quantum numbers as the Lagrangian (zero lepton number, zero baryon number, zero hypercharge,  a singlet under strong  SU(3), a singlet under weak  SU(2)).  If ${\hat {\ov \w}}^{-}_{\D ,\a \b \g}$ transforms as a pseudosuperfield with the cybersusy algebra, then so will ${\hat {\ov \w}}^{-}_{\D \; {\rm Comp},\a \b \g}$.  All such operators also give rise to the cybersusy algebra upon gauge symmetry breaking, with suitable modifications of the other spinorial pseudosuperfields,  since scalar pseudosuperfields ${\hat {\ov A}}_{\rm Composite}$   are inert.

The set of operators made in this way is clearly infinite since a product of two $
{\hat {\ov A}}_{\rm Composite}$ is another one. 

To prove that all possible operators that could be used instead of $
{\hat {\ov \w}}^{-}_{\D,\a \b \g}$ also satisfy the same cybersusy  algebra would require a complete knowledge of the cohomology, which is not available.  All that we can do now is note that the hypothesis does not seem ludicrous at this point in time.

\subsection{Other Operators}

The need for more complicated operators for the baryons is quite evident, since it is known that there are baryon resonances with spins greater than $\fr{5}{2}$.  There are at least two ways to make such operators:
\ben
\item
Add more spinors to the above operators to make simple generators with higher spin--these additional spinors would necessarily need to add up to baron number zero, so that the total operator still has baryon number one.  This requires a better understanding of the gauge cohomology.
\item
Add derivatives to the fields $\y,\oy,A,\A$ and do more cohomology to go beyond the simple generators
\een
However it does seem necessary, and sufficiently challenging, to see if the above towers of cohomology generate reasonable results using effective fields before undertaking the addional challenge of dealing with higher spins.

\section{Conclusion} 
\la{morenotationsssm}

\subsection{Remarks on the peculiar properties of the SSM from the point of view of the constraint}

The SSM, and the leptonic and hadronic solutions to the constraint equations discussed above, have the following interesting properties, as can be seen from looking at the superpotential in  Table \ref{fqwefweef1212}:
\ben
\item
The SSM has multiple indices on fields like the left quark multiplet $Q^{icp}$ and the left lepton multiplet $L^{ip}$.
\item
In the superpotential, these get contracted with sets of two  fields with  fewer indices, 
like the right quark multiplet $B_{c}^{p}$ and $T_{c}^{p}$,
and the right  lepton multiplets $P^{p}$ and $R^{p}$, multiplied by the Higgs fields $H^i$ and $K^i$. 
\item
In the superpotential,  the neutral field $J$ gets contracted with $H^i K^j \ve_{ij}$
and also for gauge symmetry breaking there is a term $J m^2$. 
\item
The generic constraint is generated by
\be
d_3  
= 
C_{\a} {\ov g}^{ijk} \A_j \A_k 
\y_{\a}^{i \dag}
\ee
plus the complex conjugate. 
\item
In the SSM this gets replaced by
\be
d_{3} =
d_{3,{\rm special}}
+
d_{3,{\rm remaining}} 
\ee
where
\be
d_{3,{\rm special}}
= \oC_{\dot \a} 
\lt \{
g \e_{ij} H^i K^j  
\oy_{J \dot \a}^{ \dag}
\ebp
+
\lt (
 p_{p q} \e_{ij}  H^j P^{ q} 
+ r_{p q} \e_{ij}  K^j R^{ q}
\rt )
\oy_{L p i \dot \a}^{ \dag}
\ebp
+
\lt (
 p_{p q} \e_{ij} L^{p i} H^j 
\rt )
\oy_{P q \dot \a}^{ \dag}
+
\lt (
 r_{p q} \e_{ij} L^{p i} K^j 
\rt )
\oy_{R q\dot \a}^{\dag}
\ebp
+
\lt (
t_{p  q} \e_{ij}   K^j T_c^{ q}
+b_{p  q} \e_{ij}   H^j B_c^{ q}
\rt )
\oy_{Q c p i\dot \a}^{ \dag}
\ebp
+
\lt (
t_{p  q} \e_{ij} Q^{c p i} K^j 
\rt )
\oy_{T q\dot \a}^{c \dag}
+
\lt (
b_{p  q} \e_{ij} Q^{c p i} H^j 
\rt )
\oy_{B q\dot \a}^{c \dag}
\rt \} 
\eb
+ {\rm Complex\; Conjugate}
\ee
\item
The terms 
\be
d_{3,{\rm remaining}} =
\ov C_{\dot \a}
\lt \{ \lt (
g \e_{ij} H^i J
+
r_{p q} \e_{ij} L^{p i}  R^{ q}
+
t_{p  q} \e_{ij} Q^{c p i}   T_c^{ q}
\rt )
\oy_{K j \dot \a}^{ \dag}
\ebp
+
\lt (
g \e_{ij}  K^j J
-
p_{p q} \e_{ij} L^{p j}   P^{ q} 
 -
b_{p  q} \e_{ij} Q^{c p j} B_c^{ q}
\rt )
\oy_{H i\dot \a}^{ \dag}
\rt \}
\ee
have not played any role in our solutions in this paper.  I suspect they play a role in the mesonic and bosonic spectrum, including the weak vector bosons, the Higgs, and the hadronic mesons, but that requires an understanding of the gauged  supersymmetry cohomology, which is presently mostly unknown.
\item
Our solutions come from
$d_{3,{\rm special}}$ and they depend on the left right assymmetry in a peculiar way, and also on the existence of the 
special term $\ov C_{\dot \a} g \e_{ij} H^i K^j  
\oy_{J \dot \a}^{ \dag}
$.
It would be interesting to see what other kinds of models, besides the SSM, allow solutions to these constraints. 
\item
From the discussion in \ci{cybersusyII}, we know that these constraints are related to the Lie Algebra of invariances of the superpotential.  But there is another thing going on in the SSM.  The solutions to the constraints that we are interested in, because they look like bound states, are not generators of the well-known invariances like $SU(3) \times SU(2)\times U(1)$ of the SSM--they are more obscure, and more interesting.

\een
 \subsection{Summary}

We have seen that the SSM provides an arena in which the constraint equations can be solved, and we have also seen that when the gauge symmetry breaks, these solutions become familiar particles for the baryons, and also for the leptons.  We have not attempted to discuss any of the observed hadronic mesons or the weak vector mesons such as the weak vector bosons, the $Z$ or the photon, or the Higgs, because they are too entangled with gauged supersymmetry.

There has been an implicit assumption that we can deal with the leptons and hadrons without worrying much about gauged supersymmetry.  Note that the simple dotspinors have no derivatives in their generators, and that the gauge particles do not have lepton or baryon number.  For these reasons, it seems likely that we can make the assumption that we have made. 

When the gauge symmetry breaks, the dotspinors are no longer invariant and they generate the cybersusy algebra that we used to break supersymmetry in \ci{cybersusyI}.  However the leptons and hadrons do not get involved with the gauge particles even in  this situation, except in the simple way that has been accounted for.

In the next paper of this series \ci{cybersusyIV}, we will calculate the propagators for the leptons with broken supersymmetry.
These were used to generate the polynomials in \ci{cybersusyI} that determine the supersymmetry breaking for the leptons, and the  spectrum of broken supersymmetry for the leptons.

Those propagators  also provide a basis for a complete analysis of the lepton spectrum, with broken supersymmetry, using cybersusy based on the SSM.

\tableofcontents

\end{document}